\documentclass[letterpaper]{jpconf}

\usepackage{graphicx}
\usepackage{amsmath}
\usepackage{epsfig}
\usepackage{rotating}

\begin{document}

%

\def\ET{\mbox{$E_T$}}
\def\pT{\mbox{$p_T$}}

\def\sqrts{\mbox{$\sqrt{s}$}}
\def\sqrtsNN{\mbox{$\sqrt{s_{\rm NN}}$}}
\def\RAA{\mbox{$R_{\rm AA}$}}
\def\TAB{\mbox{$\langle T_{\rm AB}\rangle$}}

\def\Klong{\mbox{$K_0^L$}}
\def\piplus{\mbox{$\pi^+$}}
\def\piminus{\mbox{$\pi^-$}}
\def\sqrtsNN{\mbox{$\sqrt{s_{_{\rm NN}}}$}}
\def\pizero{\mbox{$\pi^0$}}

\def\qhat{\mbox{$\hat{q}$}}


\def\Echarged{\mbox{$E_{\rm{Jet,rec}}^{\rm{Ch}}$}}
\def\Echargedcorr{\mbox{$E_{\rm{Jet,corr}}^{\rm{Ch}}$}}
\def\Eneutral{\mbox{$E_{\rm{Jet,rec}}^{N}$}}
\def\Eneutralcorr{\mbox{$E_{\rm{Jet}}^{N}$}}
\def\Ejet{\mbox{$E_{\rm{Jet,rec}}$}} 
\def\Ejetcorr{\mbox{$E_{\rm{Jet,corr}}$}}
\def\Ejetfinal{\mbox{$E_{\rm{Jet}}$}}

\def\Ecalo{\mbox{$E_{\rm{Jet,rec}}^{\rm{EMCal}}$}}
\def\Ecalocorr{\mbox{$E_{\rm{Jet}}^{\rm{EMCal}}$}}

\def\pp{\mbox{p--p}}
\def\AuAu{\mbox{Au--Au}}
\def\PbPb{\mbox{Pb--Pb}}


\def\antikT{\mbox{$\textrm{anti}-k_T$}}
\def\kT{\mbox{$k_T$}}

\def\jetpt{\mbox{$p_T^{\textrm{jet}}$}}
\def\gevc{\mbox{GeV/$c$}}
\def\gev{\mbox{GeV}}
\def\fivetev{\mbox{$\sqrt{s_{\textrm{NN}}}=5.5~TeV$}}


\title{Heavy Flavor measurements using high-\pT\ electrons in the ALICE EMCal}

\author{Mark T Heinz (for the ALICE collaboration)}

\address{Yale University, WNSL, 272 Whitney Ave, New Haven,  CT 06520, USA}

\ead{mark.heinz@yale.edu}

\begin{abstract}
Heavy flavor hadrons, i.e. those containing charm and bottom quarks, will be abundantly produced at the LHC and are important probes of the Quark-Gluon Plasma (QGP).  Of particular interest is the investigation of parton energy loss in the medium. Using heavy flavor jets we will have a pure sample of quark jets with which to study the color-charge effects on energy loss.  In addition, studies of bottom production in p+p collision at LHC energies will be utilized to further constrain the current parameters used by NLO and FONLL calculations. The talk will focus on the very high-pt electron particle identification using the EMCal detector. We present the electron reconstruction and measurements which can be achieved with 1 nominal year of Pb-Pb running at $\sqrt{s}$=5.5 TeV. We then estimate the rate of non-photonic electrons and present systematic and statistical error bars. Finally, we show preliminary results on B-jet tagging techniques in p+p which utilize jet-finding algorithms (FASTJET) in conjunction with displaced secondary vertices containing high-pt electrons.
\end{abstract}

\section{Introduction}\label{intro}

\begin{figure}[h]
\centering
\includegraphics[width=0.48\textwidth]{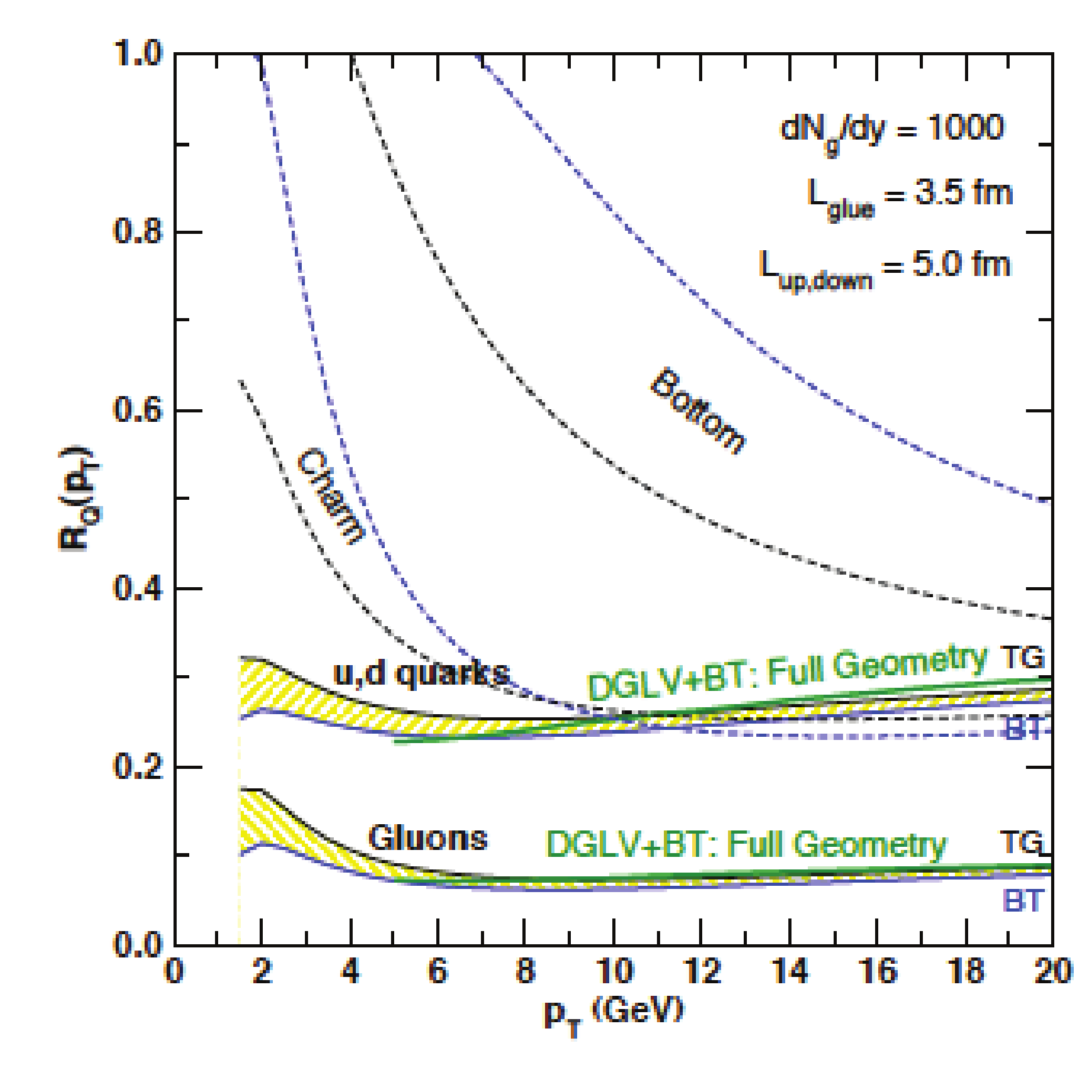}
\caption{Theory calculations of quark and gluon nuclear modification factors at RHIC 
from the DGLV model \cite{Wicks:2005gt}}
\label{fig:theory}
\end{figure}

From a heavy ion physics point of view the interest is focussed on the parton
energy loss for quarks and gluons in the strongly interacting medium produced
by the heavy ion collisions.
Heavy flavor quarks provide excellent probes since they are produced
at early times in the collisions and need to traverse the
Quark-Gluon Plasma (QGP). At intermediate \pT, the prediction 
that the energy loss of massive quarks in a colored medium
is reduced due to the suppression of forward radiation (``dead
cone effect'' \cite{Dokshitzer:2001zm}) has not been validated by
measurements at RHIC \cite{Adare:2006nq,Abelev:2006db}, which find
heavy flavor production at high \pT\ to be suppressed at the same
level as light flavor quarks and gluons. Beyond the energy range where the dead cone effect is expected to be
significant (i.e. $\ET\gg{m}$), heavy quark jet production may provide
a tool to probe the color-charge dependence of energy loss
\cite{Armesto:2005iq}. This is illustrated in Fig. \ref{fig:theory} where one 
can see that the modification, quantified by the nuclear modification factor $R_Q$, 
between charm and bottom quarks becomes comparable for $\pT > 20$ GeV/c. However
the ratio between light quarks and gluons remains constant at $\sim$9/4 due to 
the color Casimir factor.
In a more speculative vein, modelling of heavy ``quark'' propagation in a
strongly coupled fluid using the techniques of AdS/CFT has received significant 
recent attention \cite{Gubser:2009fc}, including a suggestion that the relative 
suppression of charm vs. bottom quarks is different in pQCD and AdS/CFT, with 
a magnitude that could be resolvable by experiment \cite{Horowitz:2007su}.

In addition, with the first p-p data obtained at 7 TeV basic measurements
such as  jet cross-sections, jet-profiles and jet fragmentation-functions of heavy flavor 
jets will allow us to make important tests of NLO and MLLA calculations. 

The ALICE experiment is well suited to perform the measurements of heavy flavor 
production at high \pT\ . While exclusive reconstruction of charm mesons will be carried out via the use of 
the TPC and ITS detectors, measurements of high \pT\ heavy flavor production require a fast
trigger, which is possible only for the semi-leptonic decay
mode, with a branching ratio of $\sim10\%$. The TRD has an
efficient electron trigger and good hadron rejection for $\pT<10$
GeV/c, but at higher \pT\ additional tools are required for heavy
flavor measurements. The EMCal has excellent capabilities for fast
triggering and hadron rejection, and provides unique coverage in ALICE
for heavy flavor measurements at very high \pT\ (above 10 GeV/c), and
the following studies are therefore focused on that region.

In addition the EMCal together with the central tracking system will also allow for full jet reconstruction. Jets are the primary experimental observable
for partons. Combining the jet-finding with heavy flavor tagging techniques will
allow us to obtain a reasonably clean sample of heavy flavor jets. These tagging techniques were developed 
by the Fermilab experiments (see for example  \cite{Acosta:SecVtx}). 
We will present the results from the implementation of the displaced secondary vertex method which uses heavy flavor 
electrons as "seed" particles.

\section{High \pT\ Electron Rates}\label{rates}

The goal of this section is to establish the expected \pT\ reach of heavy flavor 
hadrons in a Pb-Pb run and determine the dominant background sources. 
The primary physics sources of electrons at high \pT\ are the semi-leptonic
decays of charm (C) and bottom (B) hadrons (mostly mesons), and the decay of
W-bosons. Electrons from C-hadron decay come both from prompt
charm production and secondary decay of C-hadrons produced by the
hadronic decay of B-hadrons.

In addition, there are significant backgrounds to these sources, due both
to other physical processes and detector effects. The physics backgrounds consist
primarily of electrons from Dalitz decays of high \pT\
$\pi^0$ and $\eta$ hadrons in jet fragmentation and to a lesser extent quarkonia decays (not studied
here), while the dominant detector source are photon conversions in the ITS and beampipe which
are tracked in the TPC.

\begin{figure}[h]
\centering
\includegraphics[width=0.45\textwidth]{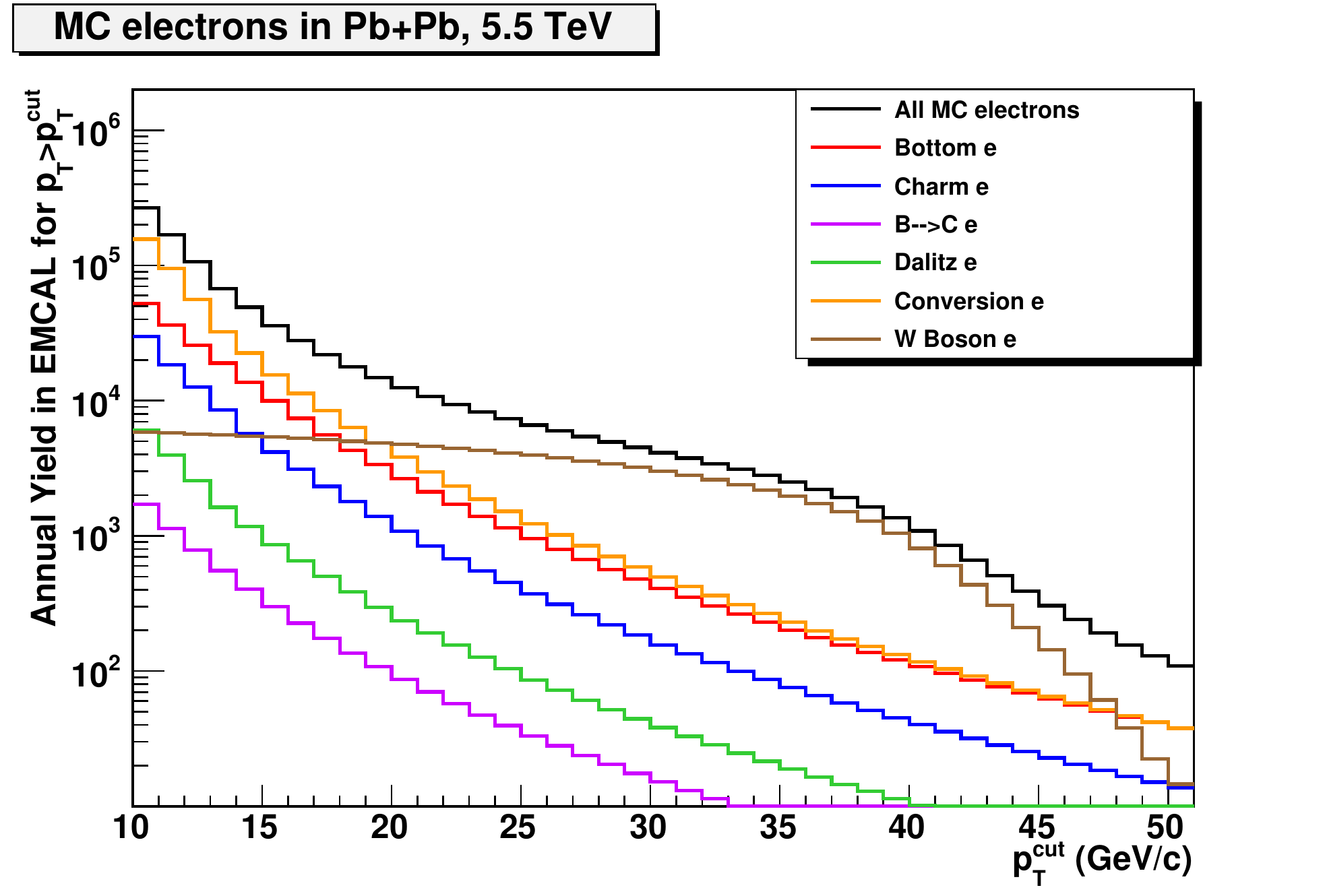}
\includegraphics[width=0.45\textwidth]{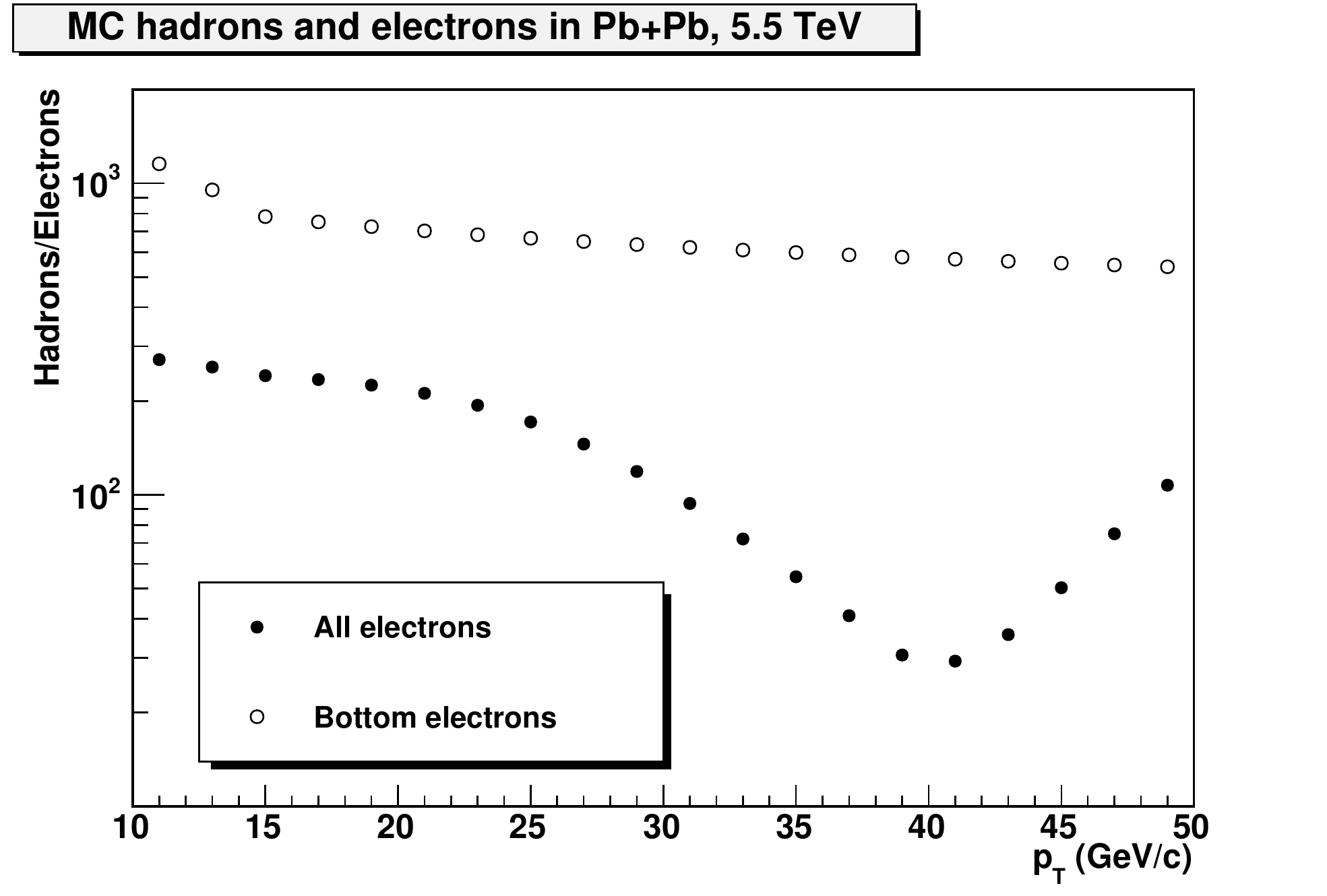}

\caption{(left) Rates of MC-Electrons from the various physics/detector sources shown as the integrated annual yield above a \pT-cut for a nominal ALICE \PbPb\ year at \fivetev. (right) Ratio of charged hadrons to electrons at the particle level expected in Pb-Pb events. The dip near 40 GeV/c is due to the contribution
of W-decay electrons.  \label{fig:rates}}
\end{figure}

To obtain the expected annual yields of single electrons in the
EMCal, a sample of PYTHIA p-p collisions at 5.5 TeV, triggered on
specific subprocesses, was simulated.  A large sample of inclusive
jet events, with one jet constrained to point towards the EMCal
acceptance, was analyzed to evaluate the dominant physics and detector
backgrounds.  A smaller sample of W-boson events was generated to
estimate the contribution from W-decay.  The signal events were
created using the ALICE standard PYTHIA heavy-flavor settings and
requiring a B-jet in the EMCal acceptance and the B-hadron to decay
semi-leptonically.

The left panel of Fig. \ref{fig:rates} shows the contributions of the
different physics sources to the distribution of electrons in
the EMCal acceptance.  Yields correspond to the expected \PbPb\
luminosity of 0.5 mb$^{-1}$s$^{-1}$ and one month (10$^6$ s) of \PbPb\
running to obtain an annual yield.

From this figure it is clear that there is a significant rate of
bottom electrons to \pT\ $\sim$ 50 GeV/c in the EMCal.  The dominant
backgrounds to heavy flavor electrons come from the photon
conversions and from W-boson decays for \pT~$>$ 20 GeV/c.  The other
significant background to measuring B-electrons will be from charged
hadrons misidentified as electrons. In order to evaluate the minimum
hadron rejection required by the electron identification algorithm,
the ratio of charged hadrons to electrons from the same simulation is
shown in the right panel of Fig.~\ref{fig:rates}.  The ratio for transverse momentum in
the range of interest from 10-50 GeV/c is a few hundred, which
sets the scale for the hadron rejection requirement.

\section{Electron Identification}\label{pid}

The method of identifying electrons in the EMCal relies on the fact
that electrons deposit all of their energy in the EMCal while hadrons
typically leave only a small fraction (i.e. MIP) of their energy in the EMCal. In order to calculate
óthe ratio of EMCal energy to reconstructed track momentum, tracks are matched to
EMCal clusters. The procedure takes a reconstructed track and extrapolates it to
the EMCal. If the distance between the extrapolated track-position
and any cluster-position is less than a given value (in this study
about the size of 1 tower) then the track is considered to be matched. In a
second loop the matches are compared and only the closest matches
are kept, thereby eliminating competing pairs. 

\begin{figure}
\centering
\begin{turn}{90}
\includegraphics[width=0.45 \textwidth]{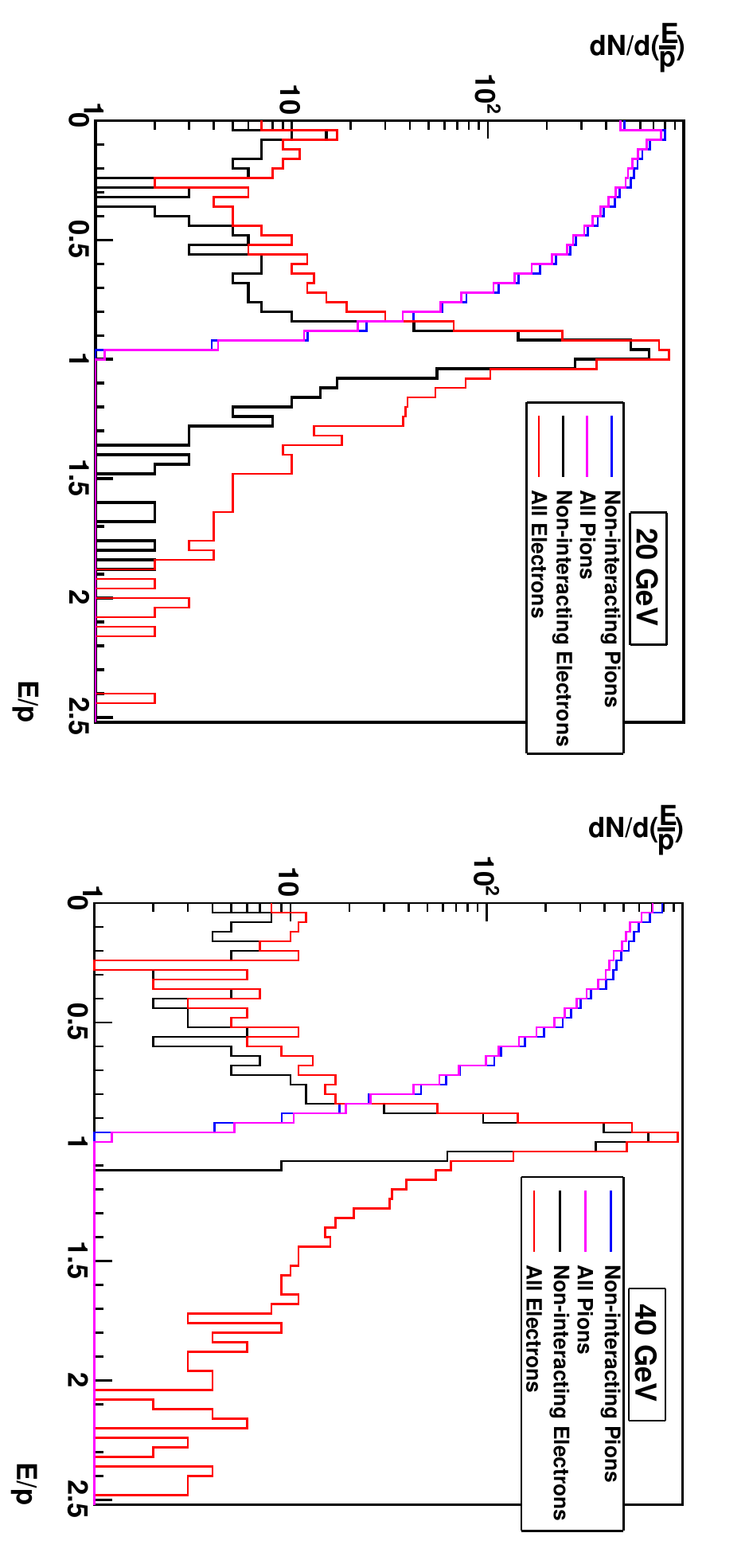}
\end{turn}
\caption{E/p distributions for single electrons and pions at momenta
  of 20 GeV/c (left) and 40 GeV/c (right). The normalization of the
  particle species relative to each other is arbitrary. The lines marked ``non-interacting''
  refer to particles that lose less than 10\% of their
  momentum before arriving at the EMCal surface.} \label{fig:pe}
\end{figure}

We use the matched cluster-track pairs, the cluster energy, and the reconstructed track momentum to calculate
the E/p ratio. This distribution is shown in Fig. \ref{fig:pe} for single electron
and pion tracks for two different momenta. The normalization of the two distributions is
arbitrary and does not reflect the respective ratio of pions to electrons. The effect of particle
 interactions with the detector material is shown for each species. This was done by comparing the MC-input to the reconstructed momentum. If the particle
had lost more than 10\% of its input momentum it was considered to have interacted
with the detector material and labeled ``interacting''.  If, on the other hand, the
reconstructed momentum was more than 90\% of the input momentum then it was considered ``non-interacting''.
Electrons are known to suffer bremsstrahlung in the detector material as can be seen in the slight
``tails" of the E/p distribution towards the larger values.


We establish an electron identification criterion by setting a lower limit on the E/p ratio.
We define the efficiency, $\epsilon$, as the number of electrons that pass
the cut divided by the total number of electrons in our sample. One
can then estimate the amount of hadron contamination by integrating
the counts above the cut value. This determines the rejection power of
the  method, defined as $\epsilon^{-1}$ for pions. The actual
electron purity in a p-p (Pb-Pb) collision will depend on the
relative ratio of pions to electrons in the data.  


\begin{figure}
\centering
\includegraphics[width=0.45 \textwidth]{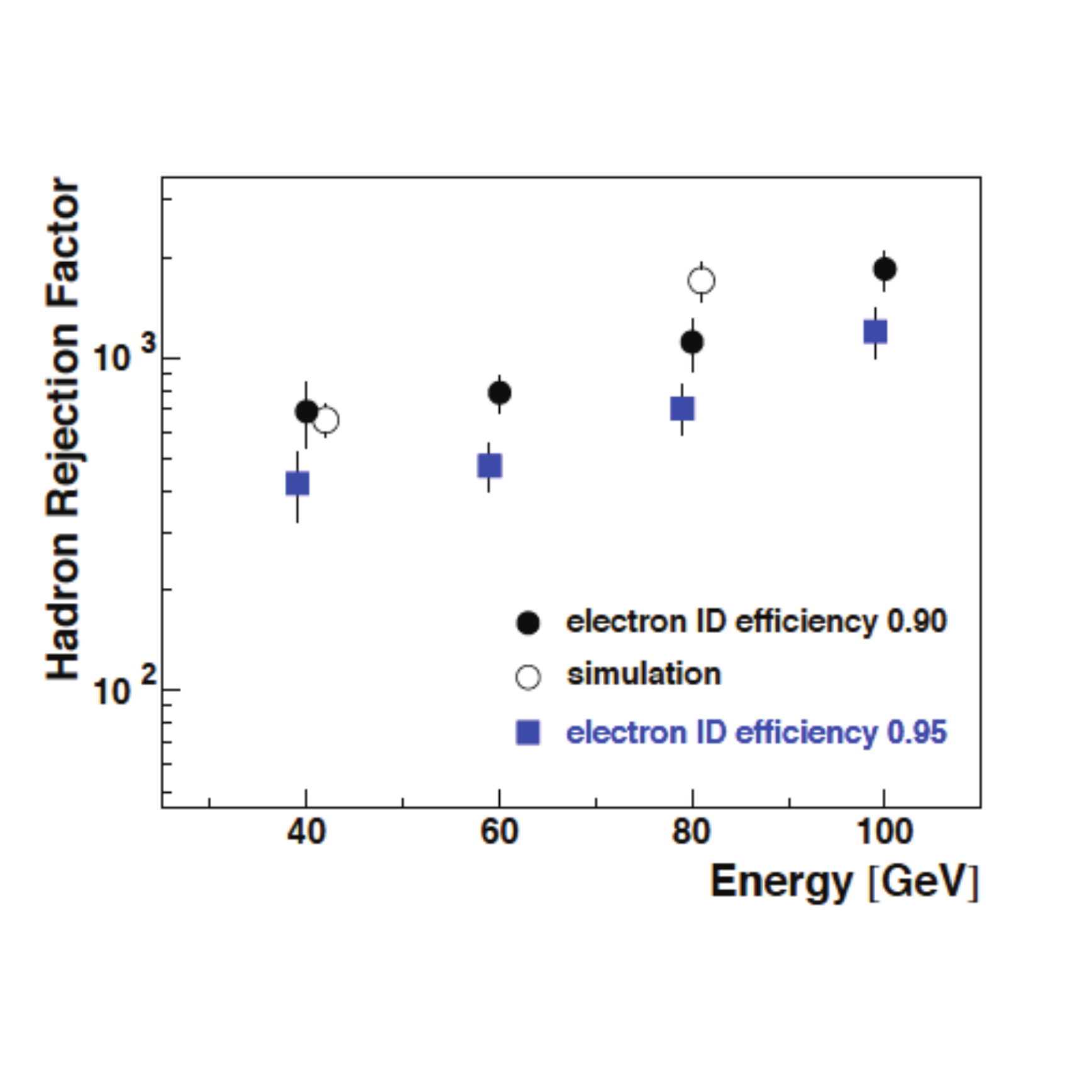}
\includegraphics[width=0.45 \textwidth]{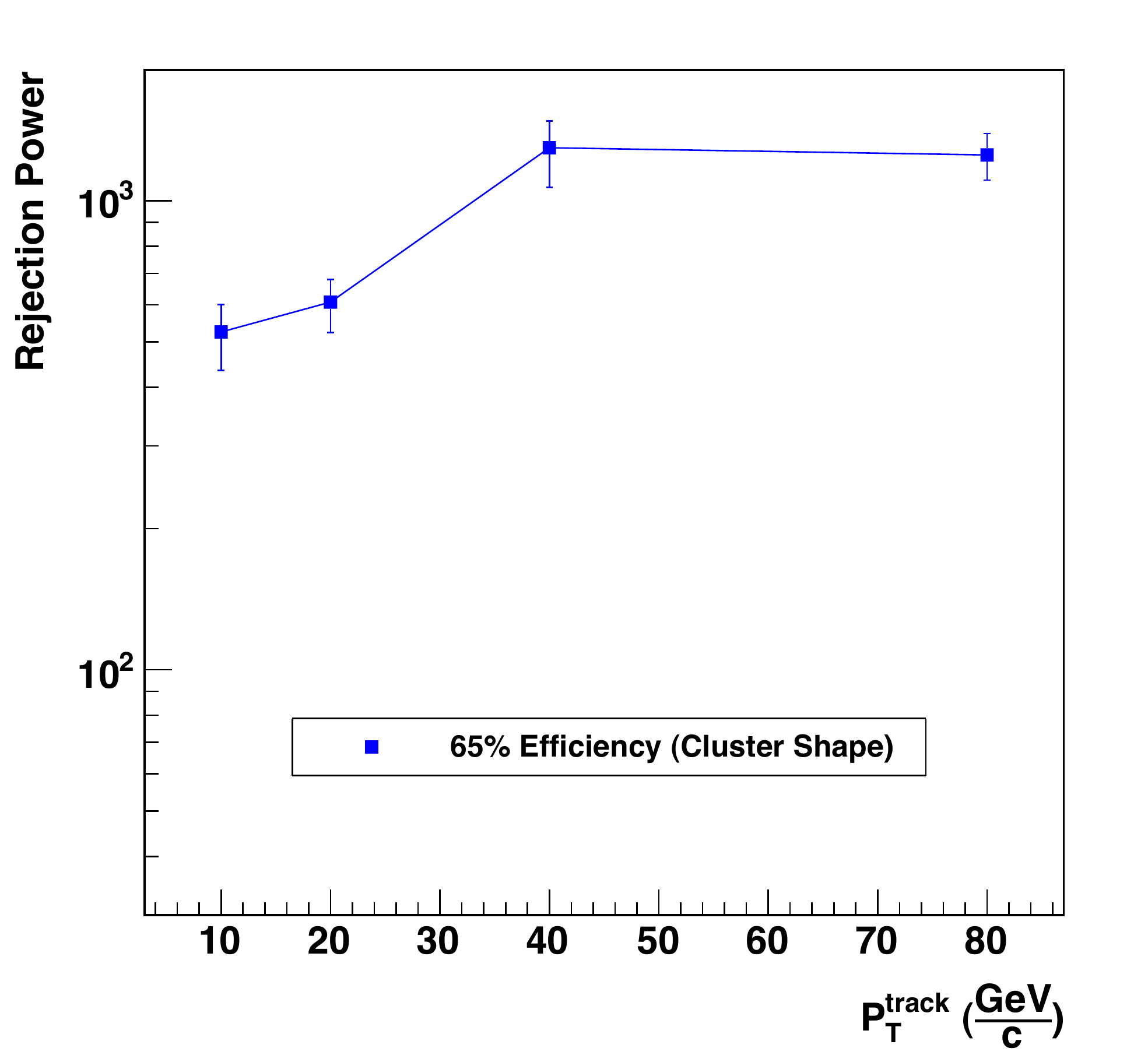}
\caption{Hadron rejection factor as a function of electron
efficiency and energy/momentum. (left) Simulation results compared to 2007 CERN test-beam data \cite{EMCAL-NIM}.
(right) Results from simulation with full detector material and optimized cuts for PYTHIA and HIJING multiplicities using 
cluster-shape and size criteria. The tracks used are required to have at least 50 TPC hits and 3 ITS hits. }
\label{fig:rejection}
\end{figure}

On the left hand side of Fig. \ref{fig:rejection} the hadron rejection power is plotted as a function of 
cluster energy and compared to the recently published 2007 CERN-SPS test-beam data  \cite{EMCAL-NIM}. The results
between simulation and real data are consistent if one considers non-interacting particles in the simulation.

The right hand side of Fig. \ref{fig:rejection} shows the results for hadron rejection power using a full GEANT
simulation of the ALICE detector material. In simulations with large hadronic and photonic backgrounds (PYTHIA, HIJING)
we determined that an additional cut on EMCal cluster shape and size was necessary to
ensure that the electron clusters are correctly identified.  This comes at the expense of electron
efficiency, which drops to $\sim$ 65\%. With these cuts the rejection power still reaches 1000 
at 40 GeV/c, yielding a resulting electron/hadron ratio of at least 3:1.
 
\section{Reconstructed Electron Spectra}\label{single}

\begin{figure}[h]
\centering
\includegraphics[width=0.45 \textwidth]{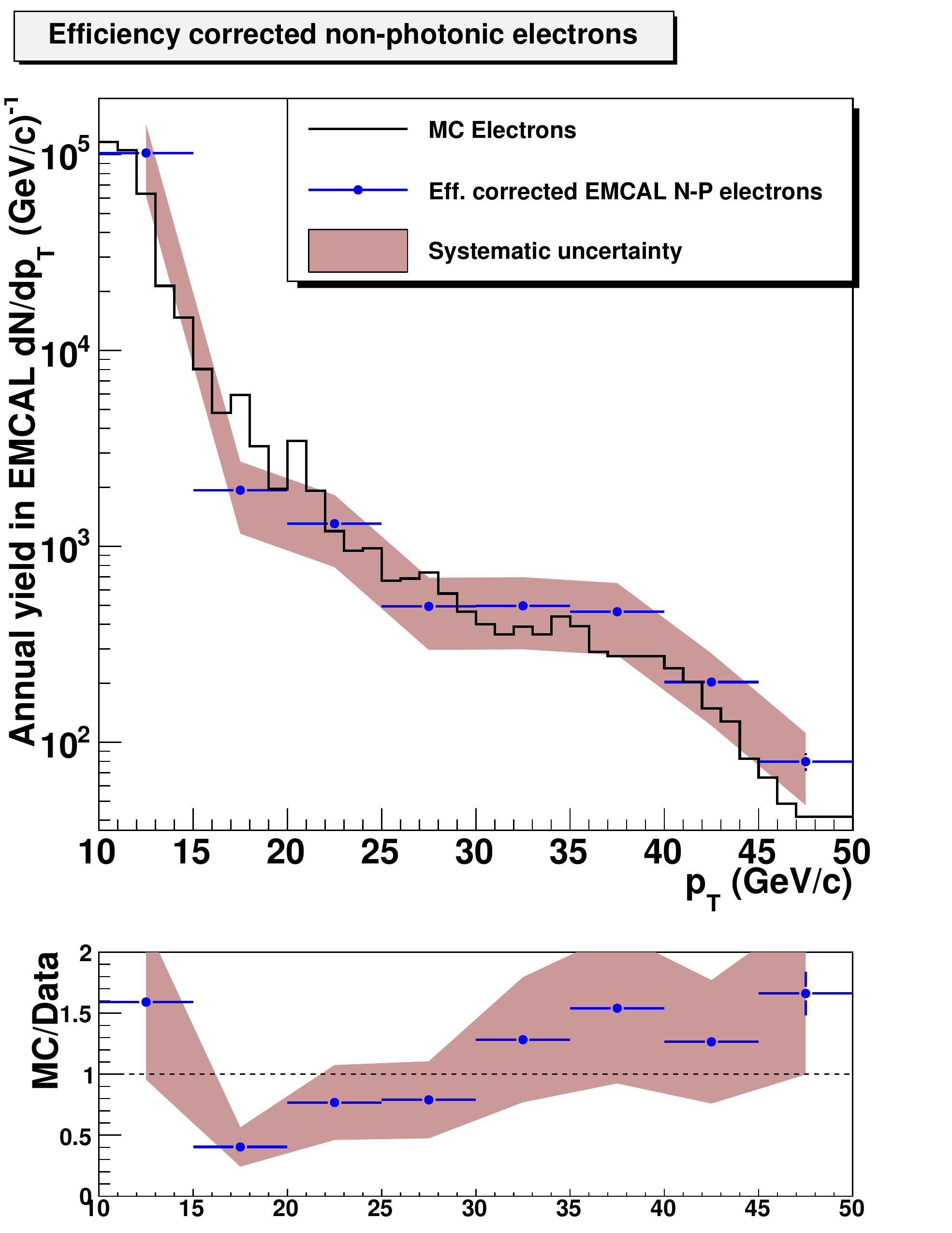}
\caption{Efficiency corrected signal of non-photonic electrons for EMCal PID compared to 
MC truth for bottom and W-decay electrons. Systematic errors due to varying the EMCal electron identification criteria are shown. }
\label{fig:npe}
\end{figure}

Utilizing the PID capabilities of the EMCal we can now reconstruct
inclusive and non-photonic electron (NPE) spectra.  In order to remove
the photonic conversion electrons from the inclusive sample, all
candidate electrons are checked against the list of reconstructed
secondary vertices (V0) created from charged particles.  Those
matching a V0 with invariant mass near that of a photon, $\rho^0$,
$\omega$ or $\phi$ hadron are considered ``photonic'' and are subtracted from the inclusive spectrum to obtain the NPE candidate yield.

To determine the total yield of non-photonic electrons, the estimated contamination from
misidentified hadrons is subtracted from the inclusive NPE candidate
spectrum and the resulting signal distribution is corrected by the
efficiency.  The result of these operations is presented in Fig.~\ref{fig:npe}.  The efficiency corrected spectrum is in agreement 
with the MC-input distribution within the systematical uncertainties. The systematic uncertainty bands were determined by 
varying the track-matching and electron identification cuts in several combinations.  The decreased efficiency resulting from tighter 
cuts is compensated by the increased purity and vice versa.  With these simulations we obtain point-by-point uncertainties that vary by less
 than a few percent about an average value of $\sim$40\%. Additional systematic effects, in particular in conjunction with the reconstruction 
 of conversion electrons, still need to be evaluated.

\section{B-Jet Tagging: Displaced Vertex Method (``DVM'')}\label{tag_yale}

This tagging method relies on the reconstruction of displaced secondary
vertices from semi-leptonic B-decays, which are typically displaced by a few hundred
$\mu$m from the primary vertex. It has been used by
CDF to identify bottom contributions in semi-leptonic muon
decays \cite{CDF:DVM}. In addition, bottom decays typically produce a large
number of charged particles by decaying
via charm mesons to lighter hadrons. The highest \pT\ hadrons are
correlated in phase-space and all point back to a common, displaced
vertex. By identifying the semi-leptonic displaced vertex consistent
with the B-meson lifetime and at least one additional hadron from
the decay we have a powerful tool to discriminate non-bottom
electrons from bottom electrons.
\begin{figure}[h]
\centering
\begin{turn}{90}
\includegraphics[width=0.5\textwidth]{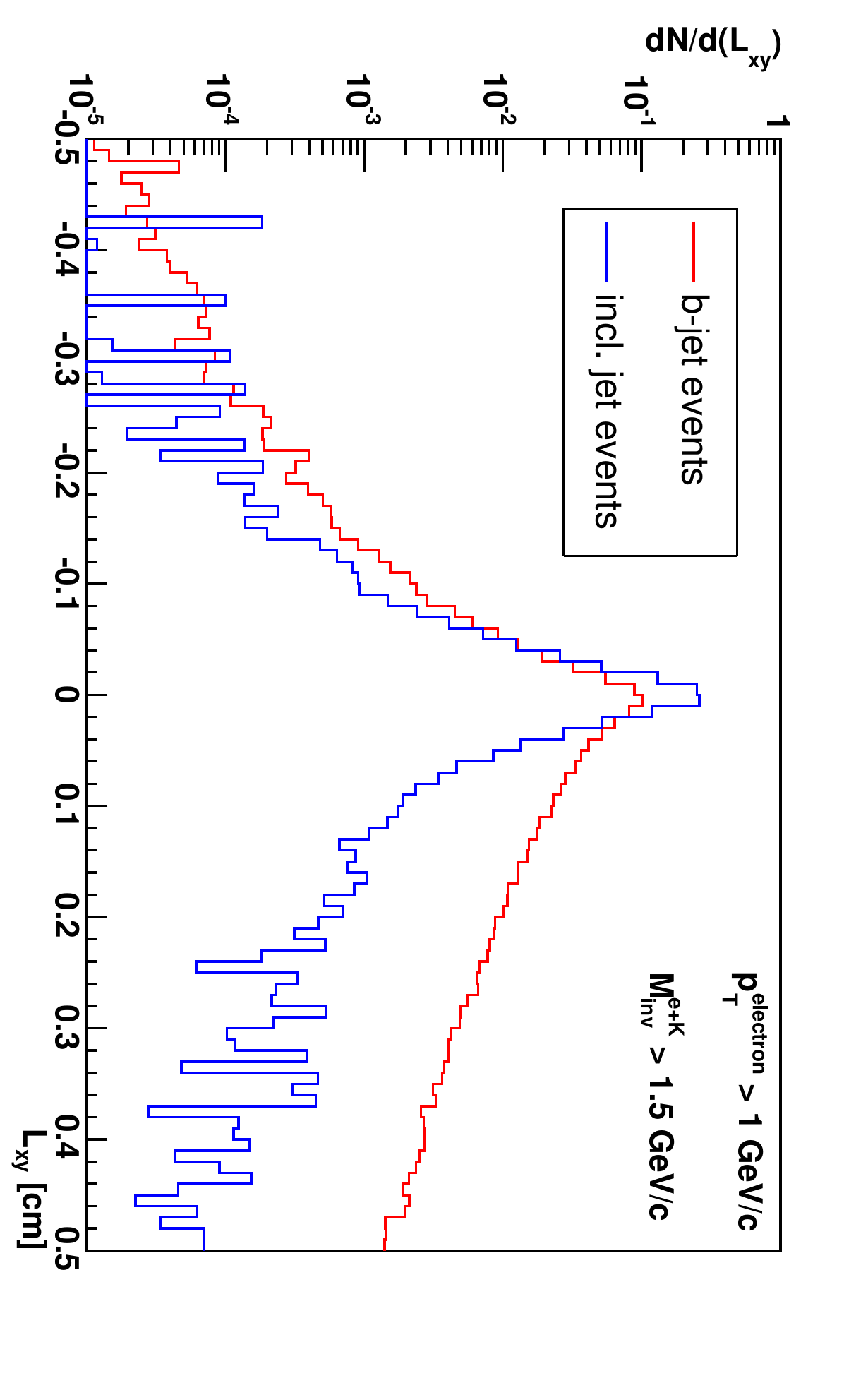}
\end{turn}
\caption{Signed DCA ($L_{xy}$) distributions from a sample of
b-jet events (``Signal'') and inclusive jet events (``Background+Signal'') at 5.5 TeV. The distributions are normalized to the same integral. The inclusive jet events also contain a small contribution of b-jets that was not subtracted. \label{fig:DVM}}
\end{figure}

The method uses a high \pT\ electron within a jet as a seed, then searches for
intermediate \pT\ hadrons (above 1.0 GeV/c) from a common secondary,
displaced vertex within a cone of $R=\sqrt{\Delta\eta^2+\Delta\phi^2} <1.0$ around the trigger.
This R value was successfully used by CDF but still needs to be optimized for the heavy
ion environment \cite{CDF:DVM}.  Electrons below \pT\ 10 GeV/c are identified using the 
combined PID of TPC, TRD and EMCAL, whereas above this cut EMCAL PID only is used
to guarantee sufficient purity of the sample. A minimum of 4 (out of 6) ITS hits are required on both
tracks to ensure sufficient spatial resolution of the secondary
vertex. Once a pair is found and its displaced vertex determined, the
quantity $L_{xy}$ (in the bending plane) is calculated:

\begin{equation}
L_{xy} = \frac{r \cdot p_{e}}{|p_{e}|} = |r| \cdot cos(\theta)
\label{Eq:Lxy}
\end{equation}

Where \textbf{r} is the vector from the primary vertex to the secondary
vertex and \textbf{p} is the electron momentum. The distribution of this quantity is
symmetric around zero for the background, but strongly biased towards
positive values for real decays. Fig. \ref{fig:DVM} shows the distributions of
$L_{xy}$ for electrons from inclusive jet (``Background+Signal'') and b-jet (``Signal'') events. More details
can be found in reference \cite{MH:WW2007}.

Based on the distribution of $L_{xy}$ we define cuts to select
electrons from b-decays. In this study we have adopted preliminary
cuts that require at least n (n=1,2,3...) tracks from secondary vertices with $L_{xy}>0.1$cm
and $M_{inv}^{e+K} > 1.5$ GeV. Cutting on the invariant mass is a
powerful discriminator to reject semi-leptonic vertices from charm. 

The next step consists of identifying such tagged electrons within jets. For this
we ran a jetfinder algorithm (FASTJET) on our data and then compared the charged jet
constituents to our collection of tagged electrons. If a jet contained a tagged
electron it was tagged as a b-jet. All jets first had to pass EMCal acceptance criteria
in $\eta$, $\phi$ and neutral energy fraction to ensure complete reconstruction
of the neutral and charged components.

\begin{figure}[h]
\centering
\includegraphics[width=0.49\textwidth]{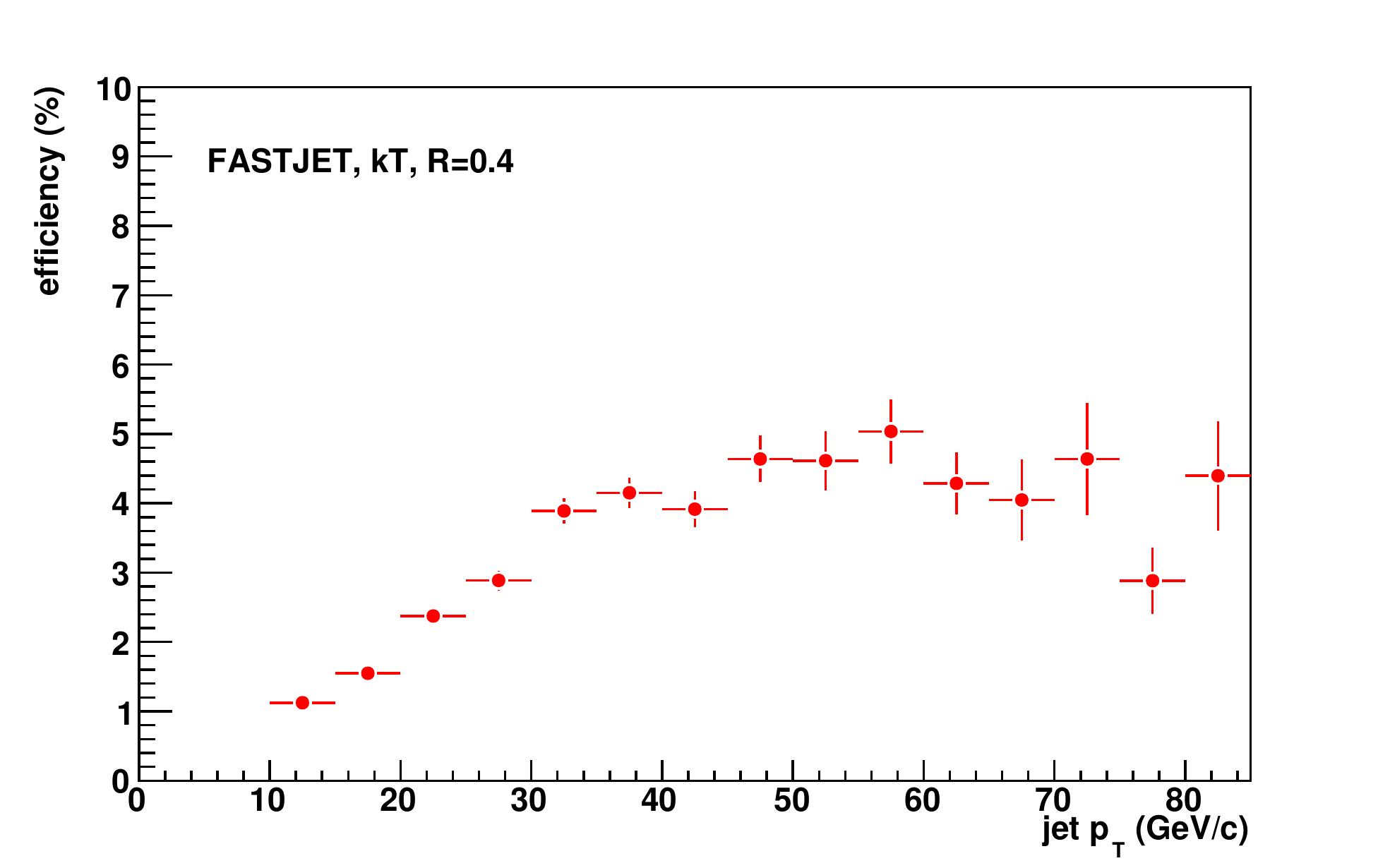}
\includegraphics[width=0.49\textwidth]{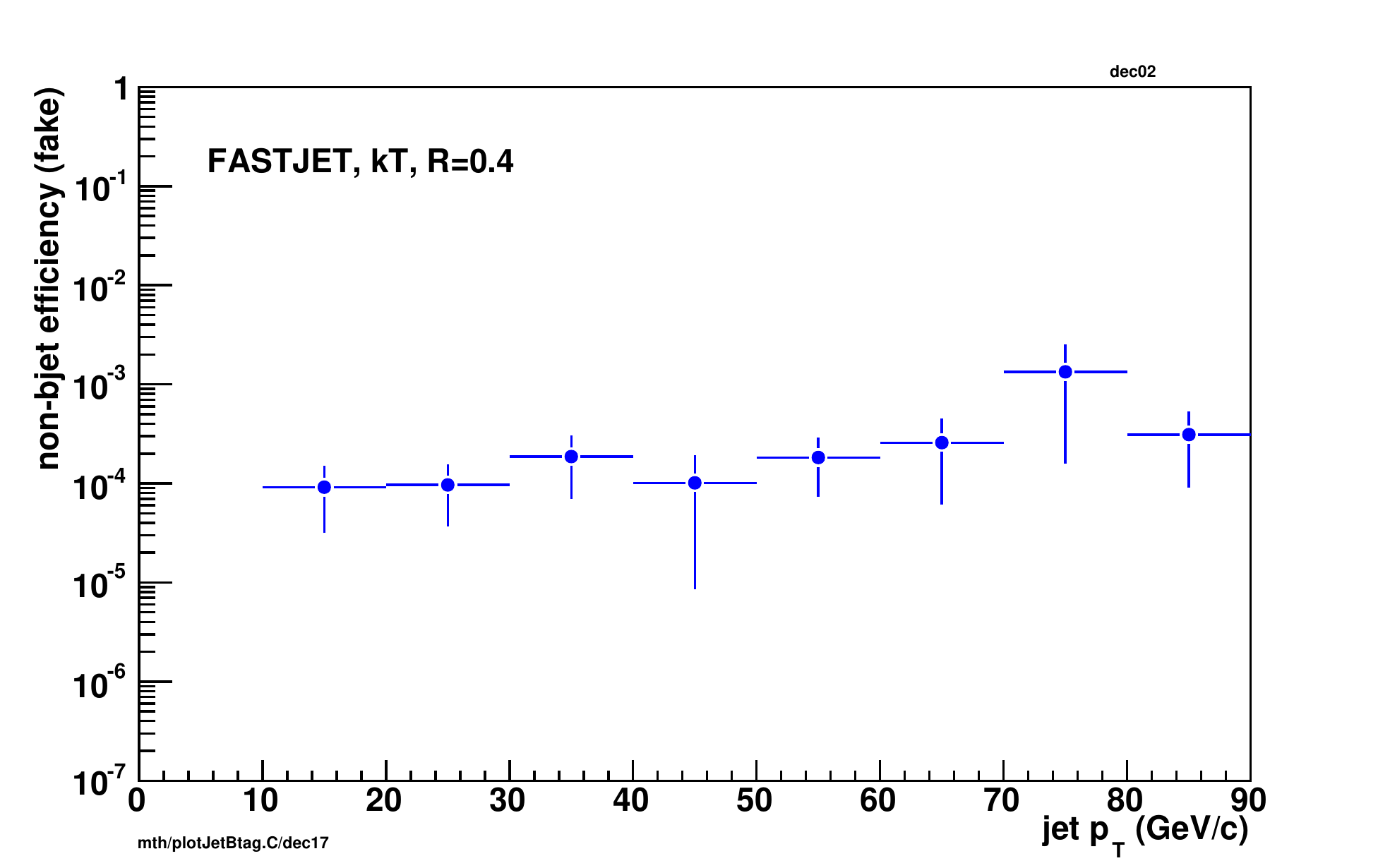}
\caption{Tagging efficiency (left) and fake rate (right) for DVM B-tagging as a function of reconstructed jet \pT . These results were obtained requiring at 
least 2 displaced vertices. Jets were reconstructed using the $k_{T}$-algorithm of the FASTJET package with R=0.4.\label{fig:jetDVMresult}}
\end{figure}

\begin{figure}[h]
\centering
\includegraphics[width=0.49\textwidth]{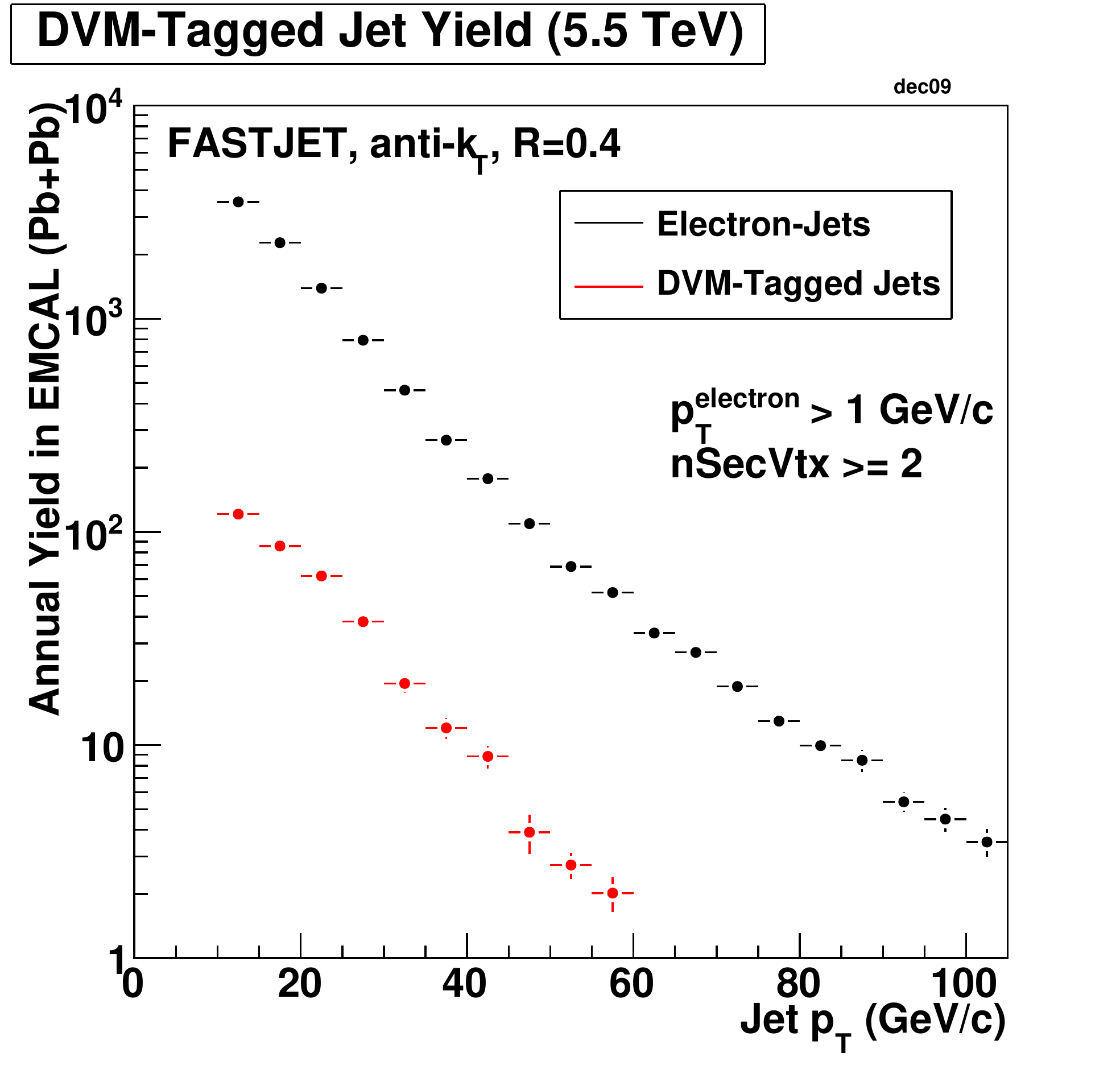}
\caption{Final yield per Pb-Pb year of B-tagged jets using the FASTJET anti-$k_{T}$-algorithm and the DVM tagging method.
\label{fig:DVMyields}}
\end{figure}

In Fig. \ref{fig:jetDVMresult} the current results of the DVM B-tagging algorithm
are shown as a function of reconstructed jet \pT. With the algorithm configured for highest purity we currently
achieve an efficiency of about 5\% for $\pT >40~GeV/c$ as compared to
MC-input B-jets. This efficiency is expected since we require one high
\pT\ electron and two hadrons from the secondary displaced vertex.
The fake tag rate, defined as the efficiency of tagging a non-B jet,
is of the order of $10^{-4}$ which means that the purity of the tagged sample, 
depending on the input model, is well above 90\%. In figure \ref{fig:DVMyields} we
show the estimated amount of tagged B-jets from one nominal year of 
Pb-Pb running which confirms that our statistical reach is up to \pT\ of $\sim$ 60 GeV/c.
These studies are still preliminary and work is ongoing to optimize
the efficiency versus fake rate of the algorithm.

\section{Conclusions}

We have shown that the EMCAL possesses excellent electron PID capabilities and extends ALICE's 
electron PID in the \pT\ range up to 80 GeV/c, with hadron rejection factors of the order of several hundreds. This capability
will allow to measure non-photonic electrons in \PbPb\ collisions for \pT\
ranges well above that previously reported at RHIC, and thus gain insight into color-charge
and/or quark mass effects of partonic energy loss. Further studies for effectively reducing conversion and
W-boson backgrounds are underway. The studies of B-tagging are at a proof-of-principle stage. We have shown
that the displaced secondary vertex algorithm yields good results in p-p collisions and allows to measure
B-jets up to a jet-\pT\ of $\sim$ 60 GeV/c.  We expect to further optimize the algorithm to find an optimal 
balance between efficiency and purity.


\end{document}